\documentstyle[12pt]{article}
\def\gev{{\rm \, Ge\kern-0.125em V}}
\begin{document}
\begin{titlepage}
\pagestyle{empty}
\baselineskip=21pt
\rightline{UMN--TH--1205/93}
\rightline{CfPA--93--th--21}
\rightline{August 1993}
\vskip .2in
\begin{center}
{\large{\bf Corrections to Bino Annihilation I: Sfermion Mixing}}
\end{center}
\vskip .1in
\begin{center}
\vskip 0.5in
{ Toby Falk,$^{1,2}$ Richard Madden,$^3$ \\
 Keith A.~Olive,$^3$  and Mark Srednicki$^{1,2}$
}\\

\vskip 0.25in
{\it

$^1${Center for Particle Astrophysics,
University of California, Berkeley, CA 94720, USA}\\

$^2${Department of Physics,
University of California, Santa Barbara, CA 93106, USA}\\

$^3${School of Physics and Astronomy,
University of Minnesota, Minneapolis, MN 55455, USA}
}\\
\vskip 0.5in
{\bf Abstract}
\end{center}
 \baselineskip=18pt

\noindent
{\newcommand{\la}{\mbox{\raisebox{-.6ex}{~$\stackrel{<}{\sim}$~}}}
{\newcommand{\ga}{\mbox{\raisebox{-.6ex}{~$\stackrel{>}{\sim}$~}}}
We consider corrections to bino annihilation due to sfermion mixing.
}}
\end{titlepage}
\baselineskip=18pt
{\newcommand{\la}{\mbox{\raisebox{-.6ex}{~$\stackrel{<}{\sim}$~}}}
{\newcommand{\ga}{\mbox{\raisebox{-.6ex}{~$\stackrel{>}{\sim}$~}}}
\def\mtw#1{m_{\widetilde #1}}
\def\tw#1{${\widetilde #1}$}
\def\beq{\begin{equation}}
\def\eeq{\end{equation}}

Within the standard model of electroweak interactions, the only plausible
dark matter candidate is the neutrino. A neutrino ($\nu_\tau$, for example)
with a mass of 10-30 eV would make a significant contribution to the
overall mass density of the Universe. A light neutrino, however, is a hot
dark matter candidate and as such cannot account for the observed structure
on small (galactic) scales. Cold dark matter alone also has its difficulties.
It is difficult to make large scale structure and conflicts with
observation of the microwave background anisotropies.  At present,
the best mix of dark matter and baryons seems to be 60-70\% cold, about 30\%
hot, and the remainder in baryons (see \cite{primack} and references
therein).  Cold dark matter still appears to be a necessity.
Before their demise, heavy (masses $\ga$ a few GeV) neutrinos were by far the
simplest cold dark matter candidates.
The minimal supersymmetric standard model (MSSM) is perhaps the simplest
logical extension which offers a cold dark matter candidate.
Because of an unbroken discrete symmetry, R-parity, the
lightest supersymmetric partner (LSP) is expected to be stable.
This makes the LSP a natural candidate
for role of the non-baryonic dark matter.

The identity of the LSP is concealed by the relatively large number of
unknown parameters of the MSSM. Nevertheless, by a
combination of i) theoretical presumptions (eg that the MSSM will
eventually be contained in a grand unified theory), ii) limits
from accelerator experiments and iii) cosmological bounds, it is
possible to identify the more plausible regions of the parameter space.
In previous studies, the lightest neutralino, a linear combination
of the supersymmetric partners of the neutral $SU(2)$ gauge boson
(neutral wino), $U(1)$ hypercharge gauge boson (bino), and the two neutral
components of the Higgs doublets (higgsinos) has been suggested as the most
likely candidate for the LSP \cite{ehnos}.
The mass matrix for these states contains
three unknown parameters: (1) $M_2$, the supersymmetry breaking $SU(2)_L$
gaugino mass; (2) $\epsilon$, the higgsino mixing mass; and (3) $\tan \beta
=v_1/v_2$, the ratio of the vacuum expectation values of the Higgs
doublets. In a large portion of this parameter space, the least massive
eigenstate is a nearly pure bino \cite{os34}.

Restricting our attention to the bino \tw B as the candidate LSP, it
is straightforward to use its couplings to the other fields in the
MSSM to compute the relic abundance. These calculations were
previously performed under a variety of assumptions: degenerate squark and
slepton masses with $\mtw q = \mtw f > \mtw B$ taken as a relatively
free parameter \cite{os34,gkt,mos};
unequal squark and slepton masses with values run down from the GUT scale using
renormalization group equations with supergravity-inspired boundary
conditions, and sfermion mixing neglected \cite{n};
the same with sfermion mixing included \cite{lny2,dn}.
Note there is also a great variety in the level of detail of these
calculations. Only in \cite{mos,dn} have all LSP--LSP annihilation channels
been included.  In \cite{dn} the annihilations of the LSP and the next lightest
neutralino (co-annihilations \cite{gs}) were also included.

In general, the sfermion mass matrix contains off diagonal terms
connecting the left and right sfermions. These are proportional
to the fermion-Higgs coupling and hence to the fermion mass.
Sfermion mixing is also dependent on the higgsino mixing mass
$\epsilon$. Since the region of the
parameter space containing the almost pure bino is in the range
of large $\epsilon$, for $m_{\widetilde B} > m_t$ we expect the
off-diagonal terms to have
a large effect on bino annihilation by exchange of a top squark.

Though mixing was included in \cite{dn}, all parameters were set
according to a particular GUT renormalization (albeit a good one).
The purpose of this paper is to examine in more detail the full effect of
including
sfermion mixing. A subsequent paper will deal with another correction
to the bino annihilation process, s-channel annihilation to
fermions via a $Z^0$ at the one loop level.

We begin by deriving the annihilation cross-section for binos,
keeping the explicit dependence on sfermion mixing.
The general form of the sfermion mass matrix is  \cite{er}
\beq
\pmatrix{ \tilde f_L^* & \tilde f_R^* } ~
\pmatrix{ M_L^2 & m^2 \cr \noalign{\medskip} m^2 & M_R^2 \cr }~
\pmatrix{ \tilde f_L \cr \noalign{\medskip} \tilde f_R }\;,
\eeq
where $m^2=m _f(A_f - \epsilon \cot \beta)$ for weak isospin +1/2
fermions and $m_f (A_f - \epsilon \tan \beta)$ for weak isospin
$-1/2$. $A_f$ is the soft supersymmetry breaking trilinear mass term
arising from superpotential terms such as $H_{1,2}Q(u^c,d^c)$ or
$H_2Le^c$, and $\epsilon$ is the Higgs mixing parameter, often written
as $-\mu$; see \cite{gh} for
an overview of couplings in the MSSM. This mass matrix is easily diagonalized
by writing the diagonal sfermion eigenstates as
\begin{eqnarray}
\tilde f_1 &=& \tilde f_L \cos\theta_f + \tilde f_R \sin\theta_f\;,\nonumber \\
\noalign{\medskip}
\tilde f_2 &=& - \tilde f_L \sin \theta_f + \tilde f_R \cos \theta_f\;.
\end{eqnarray}
With these conventions we have the diagonalizing angle and mass eigenvalues
\[
\theta_f = {\rm sign}[-m^2]\left\{ \begin{array}{ll}
{\pi\over 2} -{ 1 \over 2 } \arctan |2 m^2/
(M_L^2 - M_R^2)|,  & M_L^2 > M_R^2\;,\\
\noalign{\medskip}
{ 1 \over 2 } \arctan |2 m^2/
(M_L^2 - M_R^2)|,  & M_L^2 < M_R^2\;,
                   \end{array}
           \right.
\]
\beq
m^2_{ 1,2 } =  { 1 \over 2 }  \left[ ( M_R^2 + M_L^2 ) \mp \sqrt {
( M_R^2 - M_L^2 )^2 +
4 m^4 } \right]\;.
\eeq
Here $\theta_f$ is chosen so that $m_1$ is always lighter that $m_2$.
Note that in the special case $M_L = M_R = M$, we have
$\theta_f= $ sign[$-m^2$]$(\pi/4)$ and $ m^2_{ 1,2 } = M^2 \mp |m^2|$.

{}From here it is straightforward (if tedious) to enumerate the Feynman
diagrams contributing to the annihilation process. The dominant contribution
is due to sfermion exchange and is derived from
bino-fermion-sfermion interaction lagrangian,
\begin{eqnarray}
{\cal L}_{f \tilde f \widetilde B } &=& {\textstyle{1\over\sqrt2}}g'
\left( Y_R \bar f P_L \widetilde B \tilde f_R + Y_L \bar f P_R \widetilde B
\tilde f_L \right) + \hbox{h.c.}   \nonumber \\
\noalign{\medskip}
&=& {\textstyle{1\over\sqrt2}}g' \bar f
    \left( \tilde f_1 x P_L + \tilde f_2 w P_L +
\tilde f_1 y P_R - \tilde f_2 z P_R \right) { \widetilde B } + \hbox{h.c.}
\end{eqnarray}
where $x = Y_R \sin \theta_f$, $y = Y_L \cos \theta_f$,
$w = Y_R \cos \theta_f$, $z = Y_L \sin \theta_f$, $P_{R,L}=(1 \pm
\gamma_5)/2$, $Y_R=2 Q_f$ and $Y_L=2 (Q_f-T_{3f})$ where $Q_f$ is the
fermion charge and $T_{3f}$ is the fermion weak isospin.

To derive a thermally averaged
cross section, we make use of the technique of ref.~\cite{swo}. We expand
$\langle \sigma v_{ rel } \rangle$ in a Taylor expansion in powers of
$x = T/ m_{ \widetilde B }  $
\beq
\langle \sigma v_{ rel } \rangle = a + b x + O(x^2)
\eeq
The coefficients $a$ and $b$ are given by
\beq
a = \sum_f v_f \tilde a_f
\eeq
\beq
b = \sum_f v_f \left[ \tilde b_f + \left( -3 + {3 m_f^2 \over 4 v_f^2
    m_{\widetilde B}^2 } \right) \tilde a_f \right]
\eeq
where $ \tilde a_f $ and $ \tilde b_f $ are computed from the expansion of the
matrix element squared in powers of $p$, the incoming bino momentum, and
$v_f = (1 - m_f^2/m_{\widetilde B}^2)^{1/2}$
is a factor from the phase space integrals.

We summarize the result by quoting the computed $ \tilde a_f $:
\beq
\tilde a_f = { g'^4 \over { 128 \pi } }
  \left[ {\Delta_1 ( m_f w^2 - 2 m_{ \widetilde B } w z + m_f z^2 ) +
  \Delta_2 ( m_f x^2 + 2 m_{ \widetilde B } x y + m_f y^2 ) \over
  \Delta_1 \Delta_2} \right]^2
\label{a}
\eeq
where $\Delta_i = m_{\tilde f_i}^2+m_{\widetilde B}^2-m_f^2$.
The result for $\tilde b_f$ is too lengthy for presentation here, but
was computed and used in the numerical integrations to obtain the relic
densities. The results reduce to the results quoted in \cite{os34} in the limit
of no mixing and equal left and right sfermion masses.

We next discuss the effect of considering this
mixing interaction on the relic density and detectability of
binos. In all of the following cases we have chosen a value of
$\tan \beta=2$ and $\epsilon=\pm 3000\gev$ and assume a top quark mass
of 120 GeV. This choice of $\epsilon$ places
us in the region of parameter space where the LSP is a nearly
pure bino for a wide range of gaugino masses, $M_2 \simeq 50 - 5000\gev$.
We also take the diagonal part of the sfermion mass matrix to have a constant
value for all squarks and sleptons, and denote the common
value of $M_L$ and $M_R$ by $M$. This assumption works best when the sfermion
masses are large compared with $m_t$, $\mtw B$, and $M_Z$. For simplicity,
we take this condition to apply in all cases.  Previous
results can be viewed as a special case of this more
general form. Indeed, when mixing is neglected one must make
a particular choice of $A_f=\epsilon \cot \beta$ and $A_f=\epsilon \tan \beta$
for $T_3 =\pm 1/2$ fermions. Thus results
for the relic bino density when mixing is neglected implicitly assume
that a specific (and different) value of $A_f$ is chosen for each value
of $\tan \beta$ and $\epsilon$. In the mixed case we will also take
a common value for all $A_f$ and denote it by $A$.

Sfermion mixing has several effects on the cross-section that
can be readily understood. First, the cross-section is no longer p-wave
suppressed.  In (\ref{a}), we see that the zero temperature
cross-section now has a piece proportional to ${\mtw B}^2/\Delta^2$ rather than
$m_f^2/\Delta^2$. (Without mixing, when
$m_1 = m_2$, this piece disappears.) For $\mtw B > m_f$, this serves to
increase the cross-section.
However this aspect of mixing produces little effect on $b$, which already
includes terms proportional to ${\mtw B}^2/\Delta^2$.  On the other hand,
the splitting of the sfermion masses makes $\Delta_2 > \Delta_1$.
The cross-section (relative to the unmixed case) may
be either larger or smaller depending on which mass term is held fixed.
When the diagonal elements are fixed, the presence of the lighter sfermion
in the mixed case tends to increase the cross-section;  however, since the
bino couples with different strengths to the two sfermion mass eigenstates,
the cross-section can actually decrease [see fig.~(1)].
When the lightest mass eigenvalue $m_1$ is fixed,
the cross-section goes down as half the sfermions get heavier.

In fig.~(1), we show the effect of mixing on the coefficient
$a$ given by (\ref{a}) as a function of the bino mass (which
for our chosen values of $\epsilon$ is directly proportional to the
gaugino mass $M_2$).
In the upper and lower curve where mixing is included,
we have taken $A_f=0$ and chosen $M$ so that $m_1=m_{\widetilde B}$,
the smallest mass consistent
with the identification of the bino as the LSP.
In the upper curve $\epsilon=3000\gev$, while in the lower curve
$\epsilon=-3000\gev$. The middle
curve, where mixing is neglected, has the same value of $M$
as the other curves but
with off diagonal mass entries set to zero by adjusting the $A_f$'s
 to the appropriate values.
In these extreme cases, mixing may either increase or decrease
the cross-section depending on the sign of $\epsilon$.
The increase in cross-section is primarily due to the presence
of a lighter sfermion in the propagator. When $\epsilon=-3000\gev$,
there is a cancellation which then decreases the cross-section
at larger $m_{\widetilde B}$. (However, in this case the
relic abundance turns out to be too high, as we will discuss
below.)
Notice also the large effect of crossing the top quark
threshold.

In fig.~(2), we fix the relic density $\Omega_{\widetilde B} h^2$ and
consider the effects of mixing on the zero temperature cross-section.
$\Omega_{\widetilde B}$ is the mass density of the binos in units
of the critical density, and $h$ is the Hubble parameter in units of
100 km s$^{-1}$ Mpc$^{-1}$.  In the mixed case, we set $A_f = 0$, and
for each value of $\mtw B$, vary $M$ until the value of $\Omega_{\widetilde B}
h^2$ equals $1/4$, corresponding to a critical mass density of binos with
a value of $h = 1/2$.
Here $\epsilon=3000\gev$ only, as there are no solutions to
$\Omega_{\widetilde B} h^2=1/4$ with $\epsilon=-3000\gev$.
In the unmixed case, we set the $A_f$'s appropriately
and again adjust $M$ so that  $\Omega_{\widetilde B} h^2 = 1/4$.  The ratio
of the $a$ values in the mixed and unmixed cases is plotted in fig.~(2).
Notice here that $a$ is approximately twice as large in the mixed
case.

The enhanced value for the zero-temperature cross-section has
implications for the prospects of detecting neutralino
dark matter. Among the proposed effects signaling the presence of
the particles
comprising dark matter are (i) gamma ray line flux from annihilations
in the galactic halo \cite{gamma} (ii) high energy neutrino flux from
dark matter particles annihilating after being trapped in the Sun
or the Earth \cite{sos} and (iii) direct cryogenic detection
of DM particles scattering off nuclei \cite{gw} (for a review,
see \cite{cryo}). Referring
again to fig.~(2) we see that the inclusion of mixing allows the
relic abundance of binos to be the same as in the nonmixing case
in spite of the increased cross section $a$, due to a large decrease in the
p-wave part of the cross-section given by $b$.   The increased value of $a$
will directly increase the expected gamma
flux in (i) as the flux is proportional to $a$.  Bino annihilation in the
sun is not affected, as the rate there is limited by the sun's ability to trap
binos. In cryogenic detectors, the effect will also be important,
as pointed out in \cite{sw}, due to the effect of mixing on the elastic
cross-section (see also \cite{fos,dn2} for more on the elastic scattering
of binos).

In fig.~(3), we show the region of the $\mtw B - A$ parameter
space excluded by the conditions 1) $\Omega_{\widetilde B} h^2 < 1/4$,
$\,$ 2) the lightest sfermion is heavier than the bino, and 3) the lightest
sfermion is heavier than $74\gev$.  Here we have set $\epsilon=3000\gev$, but
note that for other choices of $\epsilon$ the only substantial change
in the picture is a shift in the values of $A_f$ (since the stop is
dominant over much of the space, the relic abundance depends largely
only on the off-diagonal matrix elements for the stops). The upper limit
on the bino mass has the same physical origin as in the unmixed case.  For
no mixing, the smallest value of $\Omega_{\widetilde B}$
occurs when $m_{\widetilde B}
= m_{\widetilde f}$ ; as both  $m_{\widetilde B}$ and  $m_{\widetilde f}$
are increased, $\Omega_{\widetilde B}$ also increases.
When we include mixing, the
upper limit depends on the trilinear mass term $A$, as shown in fig.~(3).
Again, the smallest value of $\Omega_{\widetilde B}$
for a fixed $A_f$ and  $m_{\widetilde B}$
occurs when the bino is as heavy as the lightest sfermion.

The lower limit on  $m_{\widetilde B}$ seems to be a general feature of
including mixing.
The stop mixing effect becomes large outside of a region a few hundred
GeV wide near $A_f = 1500\gev$ and the condition that $m_{\widetilde B}
> m_{\widetilde t}\;$ forces a large value for the diagonal entries, $M$.
The large sfermion masses suppress the annihilation rate below the top
threshhold.  Near $A = 1500\gev$, there is little stop mixing, $M$ can be
much smaller, and binos can
efficiently annihilate into $\tau$'s and $b$'s.  In this region,
the condition $m_{\widetilde f} > 74\gev$ limits the annihilation rate and
determines the lower bound on $m_{\widetilde B}$.

Finally, in fig.~(4) we show the effect of $A$ in the off-diagonal mass term.
The bino mass is fixed at $150\gev$, $\epsilon=3000\gev$ and
we plot $\Omega_{\widetilde B} h^2$ as a function of $A$ for several values
of $M$. We effectively pass through the nonmixing case at
a value of $A=1500\gev$ (where the stops become unmixed).
Again the points plotted are the
largest range of $A$ consistent with the identification of
the bino as the LSP. In the absence of mixing, as the sfermion mass
was increased the cross section decreased leading to increased
$\Omega_{\widetilde B} h^2$. So choosing a desired value for
$\Omega_{\widetilde B} h^2$ fixed
the value of the sfermion mass in relation to the bino mass.
Now with the addition of the parameter $A$ it is possible
to achieve a range of possible sfermion masses consistent
with a given relic density as is evidenced in fig.~4.

To summarize, we have calculated the effect of right and left sfermion mixing
on the annihilation cross-section for
binos.  For a fixed galactic halo density of binos, we find mixing enhances the
zero-temperature cross-section; this may improve prospects of detecting
neutralino dark matter through gamma ray flux from annihilations in
the halo or cryogenic detection of binos scattering from nuclei.
When the mixing effect for stops becomes pronounced we also find
stronger upper and lower bounds on $\mtw B$
coming from the requirement that
$\Omega_{\widetilde B} h^2 < 1/4 \;$ and the consistency condition
that the bino be the LSP.
Finally, given a halo density $\Omega_{\widetilde B} h^2$,
bino mass $m_{\widetilde B}$ and choice of $\epsilon$ we note
that in the mixing case we are allowed a range of sfermion masses
through variation of the trilinear mass terms $A_f$, whereas
the sfermion mass was uniquely determined in the unmixed case.

\bigskip

\vbox{
\noindent{ {\bf Acknowledgements} } \\
\noindent  This work was supported in
part by DOE grant DE--AC02--83ER--40105, and NSF grants PHY--91--16964 and
AST--91--20005.  The work of KAO was in addition
supported by a Presidential Young Investigator Award.
}}
}

\newpage
\noindent{\bf{Figure Captions}}

\vskip.3truein

\begin{itemize}

\item[]
\begin{enumerate}
\item[]
\begin{enumerate}
\item[{\bf Figure 1:}] The zero-temperature cross-section, $a$, as a function
of the bino mass when the diagonal entries of the sfermion mass matrix are
held fixed.
The upper curve ($\epsilon=3000\gev$) and lower curve ($\epsilon=-3000\gev$)
assume sfermion mixing. In the middle dotted curve the values of
$A_f$ are taken in such a way as to create no mixing, but with the
same diagonal mass entries as in the mixed cases.

\item[{\bf Figure 2:}] The ratio of the zero-temperature cross-sections,
$a_{mix}/a_{nomix}$
as a function of the bino mass when the relic density
$\Omega_{\widetilde B} h^2 = {1 \over 4}$
is fixed. $\epsilon=3000\gev$ and the diagonal mass choices are made as in
Figure 1.

\item[{\bf  Figure 3:}] Allowed regions
($\Omega_{\widetilde B} h^2 \leq {1 \over 4}$)
in the $m_{\widetilde B} - A_f$ plane for $\epsilon=3000\gev$.

\item[{\bf  Figure 4:}]  The relic bino density $\Omega_{\widetilde B} h^2$
at $\epsilon=3000\gev$ as a function of $A_f$ for several values (as labeled)
of the diagonal entry of the sfermion
mass matrix, $M$.

\end{enumerate}
\end{enumerate}
\end{itemize}
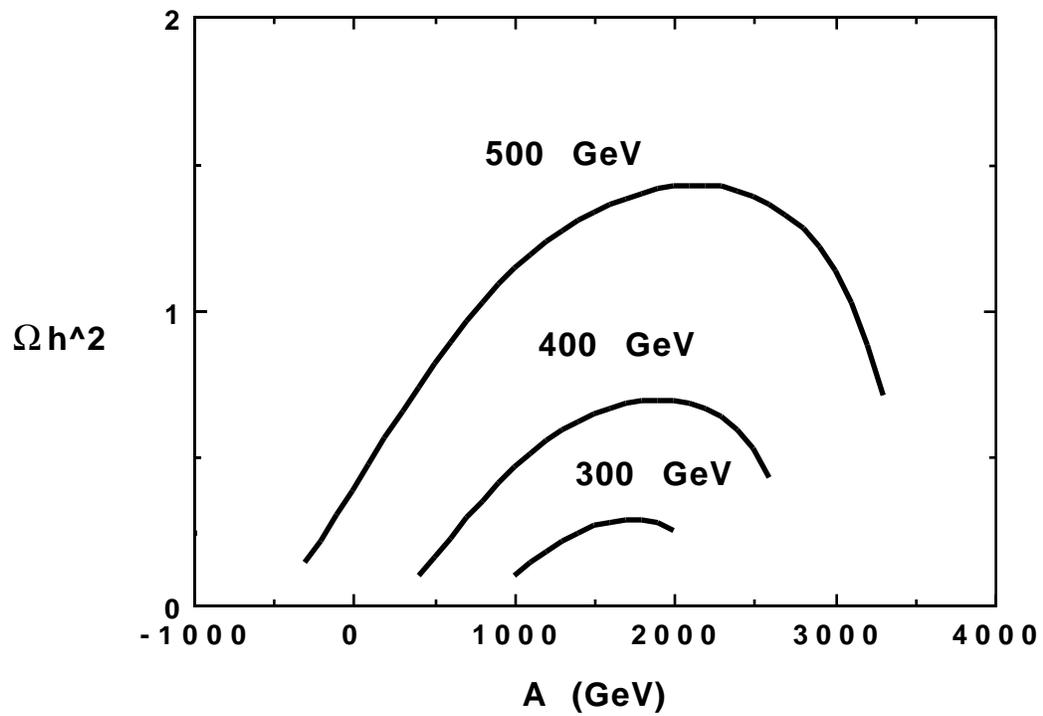
\end{document}